\documentclass{PoS}

\newcommand{\rmin}{r_{\mathrm{min}}}
\newcommand{\rmax}{r_{\mathrm{max}}}
\newcommand{\msbar}{\overline{MS}}
\usepackage{subfigure}
\usepackage{graphicx}

\title{The spatial string tension and dimensional reduction in QCD}

\ShortTitle{The spatial string tension and dimensional reduction in QCD}

\author{\speaker{Jack Liddle}, for the RBC-Bielefeld collaboration.\\%
Fakult\"{a}t f\"{u}r Physik, Universit\"{a}t Bielefeld, Universit\"{a}tsstrasse, D-33615 Bielefeld, Germany \\
       E-mail: \email{liddle@physik.uni-bielefeld.de}}

\abstract{The spatial string tension for $3+1$ dimensional QCD at finite temperature is measured. The gauge configurations were generated with two light and one heavier strange quark on lattices of size $16^3 4$ and $24^3 6$. This spatial string tension is compared with the string tension of the 3 dimensional pure gauge theory together with the temperature dependent 2-loop running coupling. Further comparison is made with predictions from dimensionally reduced effective theories.}

\FullConference{The XXV International Symposium on Lattice Field Theory\\
                July 30 - August 4 2007\\
                Regensburg, Germany}

\begin{document}

\section{Introduction}

The concept of dimensional reduction can be used to relate the properties of $3+1$ dimensional QCD at high temperatures to the properties of a $3$ dimensional pure gauge theory.  We wish to make a comparison with theoretical predictions by performing a Monte Carlo study.  The quantity of interest will be the spatial string tension.

\section{Dimensional reduction}
At high temperatures the QCD action can be reduced to a 3 dimensional pure gauge theory~\cite{Appelquist:1981vg,Linde:1980ts}.  The QCD action is written,
\begin{equation}
S_{\mathrm{QCD}} = \int^{\frac{1}{T}}_0 dt \int d^3 x \left (\frac{1}{g^2} F^{\mu\nu} F_{\mu\nu} + \overline{\psi} \gamma.D \psi\right ).
\end{equation}
The masses of modes are given by $2 \pi n T$ for the bosonic fields and $2 \pi (n+1) T$ for the fermionic fields.  At high temperatures only the zero mode of the bosonic field remains massless.  If one integrates out the massive modes then one is left with an effective theory for length scales $>> \frac{1}{T}$, called Electric QCD with an action

\begin{equation}
S_\mathrm{EQCD} = \int d^3 x \frac{1}{g_E^2} \mathrm{Tr} F_{ij}(x) F_{ij}(x) + \mathrm{Tr} [D_i,A_0(x)]^2 + m^2_E \mathrm{Tr} A_0(x)^2,
\end{equation}
with an adjoint Higgs field, $A_0$.  The electric coupling, $g_E^2$, is related to the running coupling $g^2(T)$ of $3+1$ dimensional QCD,
\begin{equation}
\label{eqn:0loopcoupling_electric}
g_E^2 = g^2(T) T,
\end{equation}
and the $A_0$ field has an electric mass,
\begin{equation}
m^2_E \simeq g^2(T) T^2.
\end{equation}
If one further integrates out the $A_0$ field from EQCD we are left with a pure gauge theory in $3$ dimensions, which is an effective theory for length scales $>> \frac{1}{gT}$.  This theory is known as Magnetic QCD and has the action
\begin{equation}
	S_\mathrm{MQCD} = \int d^3 x \frac{1}{g_M^2} \mathrm{Tr} F_{ij}(x) F_{ij}(x).
\end{equation}

\section{Lattice Setup}
The study was performed on lattices of size $16^3 4$ and $24 ^3 6$ (a single lattice of size $32^3 6$ was also studied).  The gauge sector uses a Symanzik improved action involving $1\times2$ plaquettes.  The fermionic sector uses a p4fat3 action to improve rotational and flavour symmetry, with $N_f = 2 + 1$ flavours of fermions.  The strange quark mass, $m_s$, is set so that $m_s \approx m_s^{\mathrm{phys}}$.  The configurations were generated along the line of constant physics defined by $m_\pi \approx 220 MeV$ and $m_K \approx 500 MeV$.  The updates were performed using an RHMC algorithm with $\it{O(10^4)}$ trajectories for each combination of coupling and quark masses studied~\cite{Janvdh}.

\begin{figure}[ht]
\begin{center}
\subfigure{\includegraphics[angle=-90,totalheight=5.2cm]{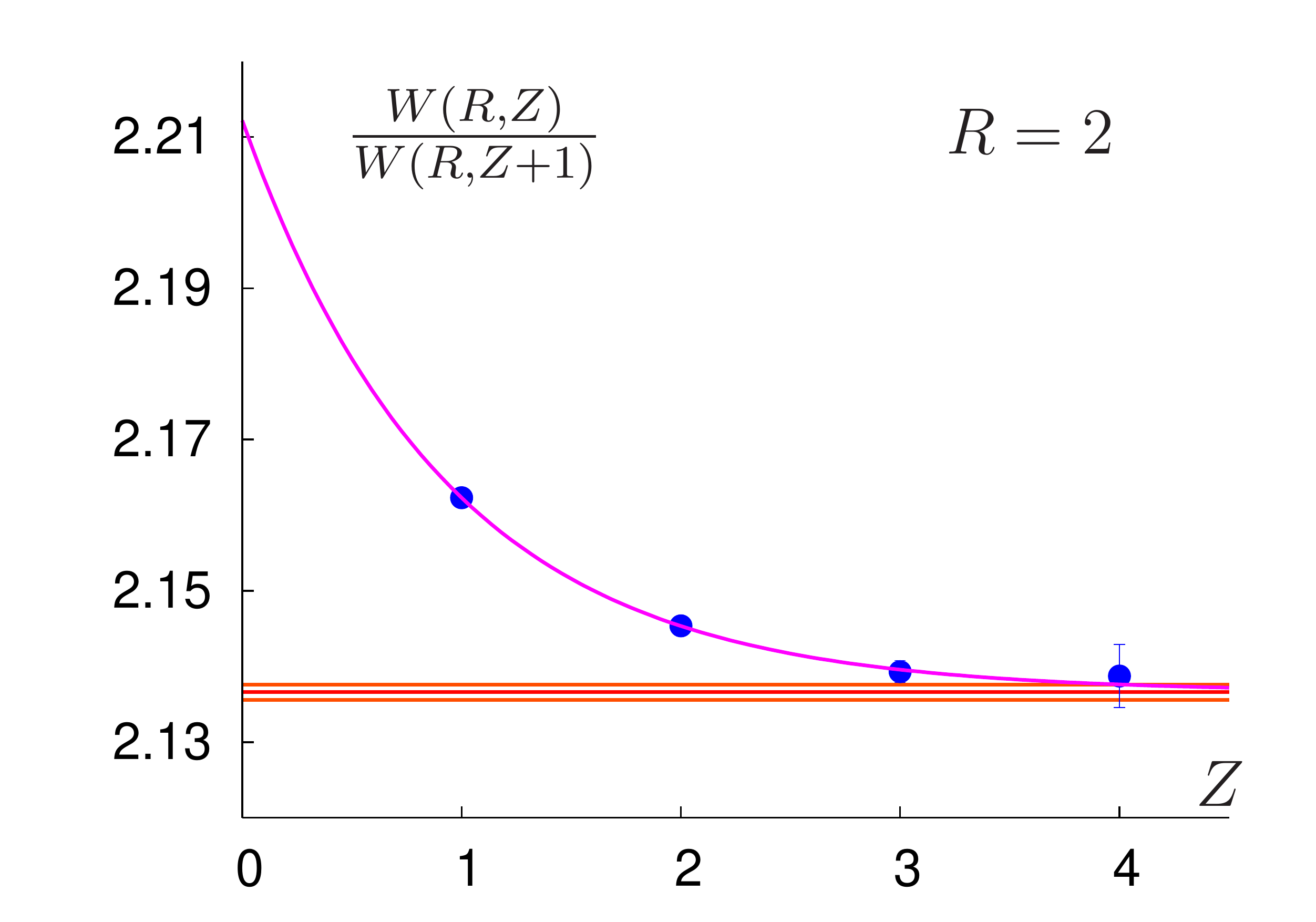}}
\subfigure{\includegraphics[angle=-90,totalheight=5.2cm]{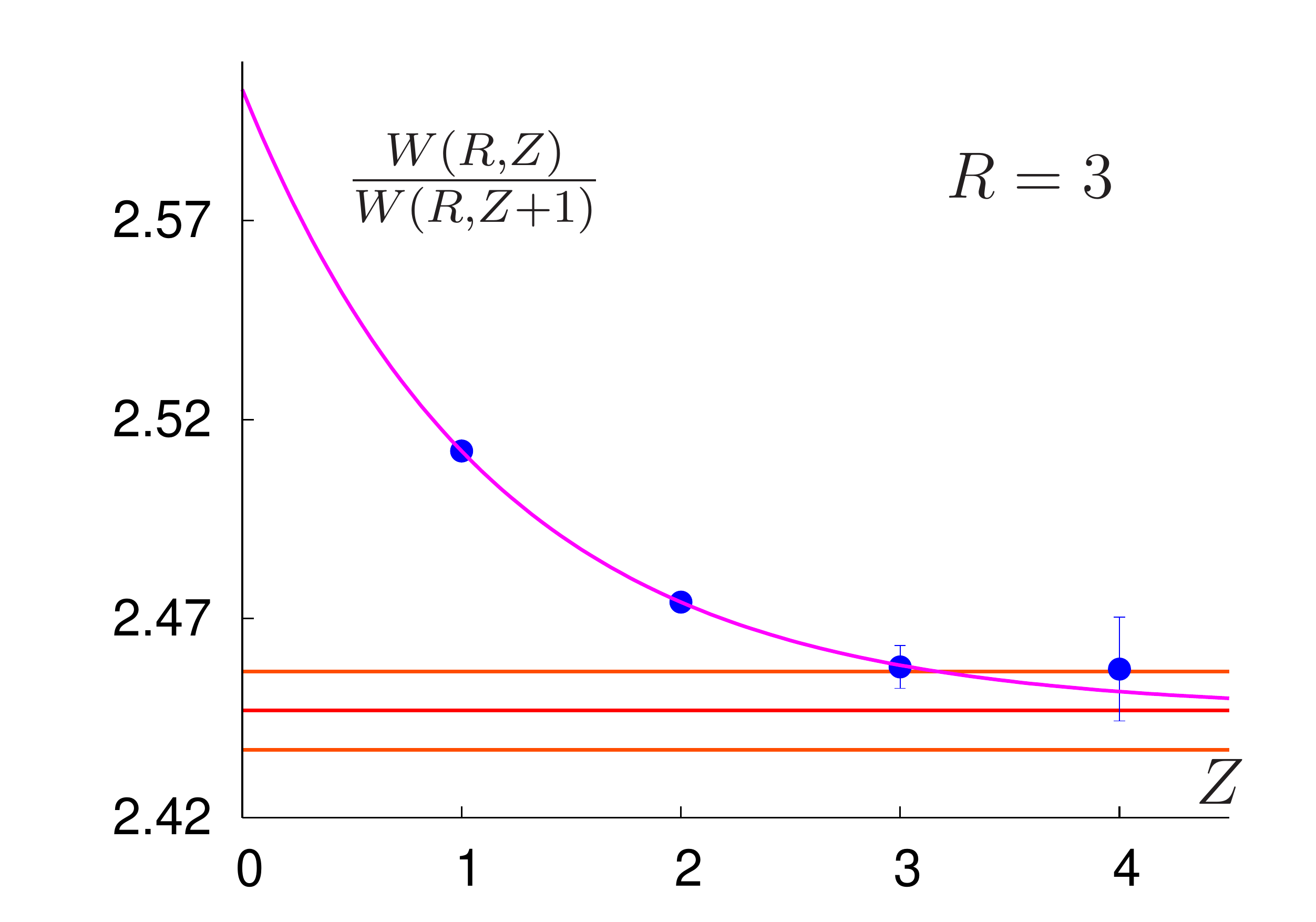}} \\
\subfigure{\includegraphics[angle=-90,totalheight=5.2cm]{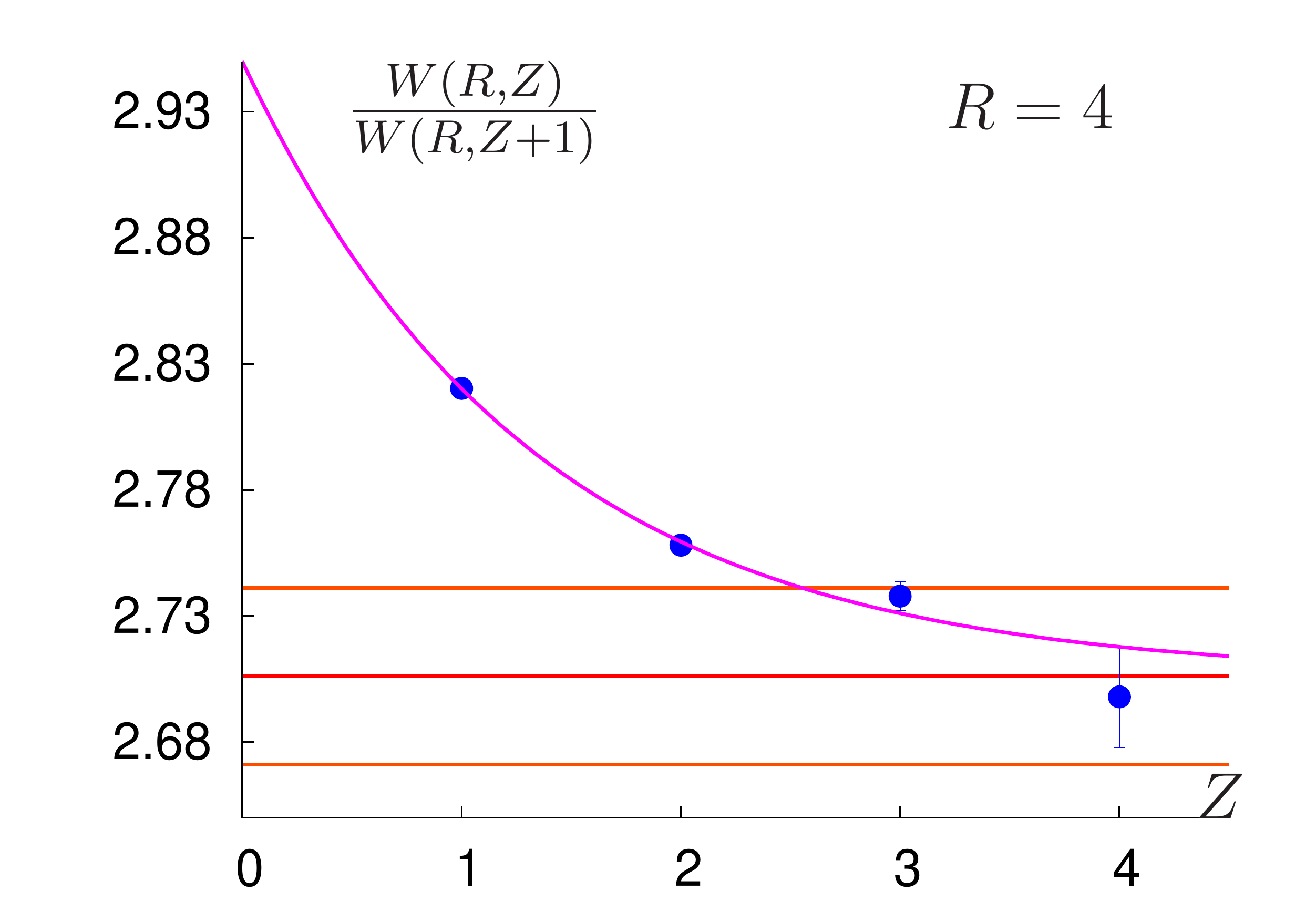}}
\subfigure{\includegraphics[angle=-90,totalheight=5.2cm]{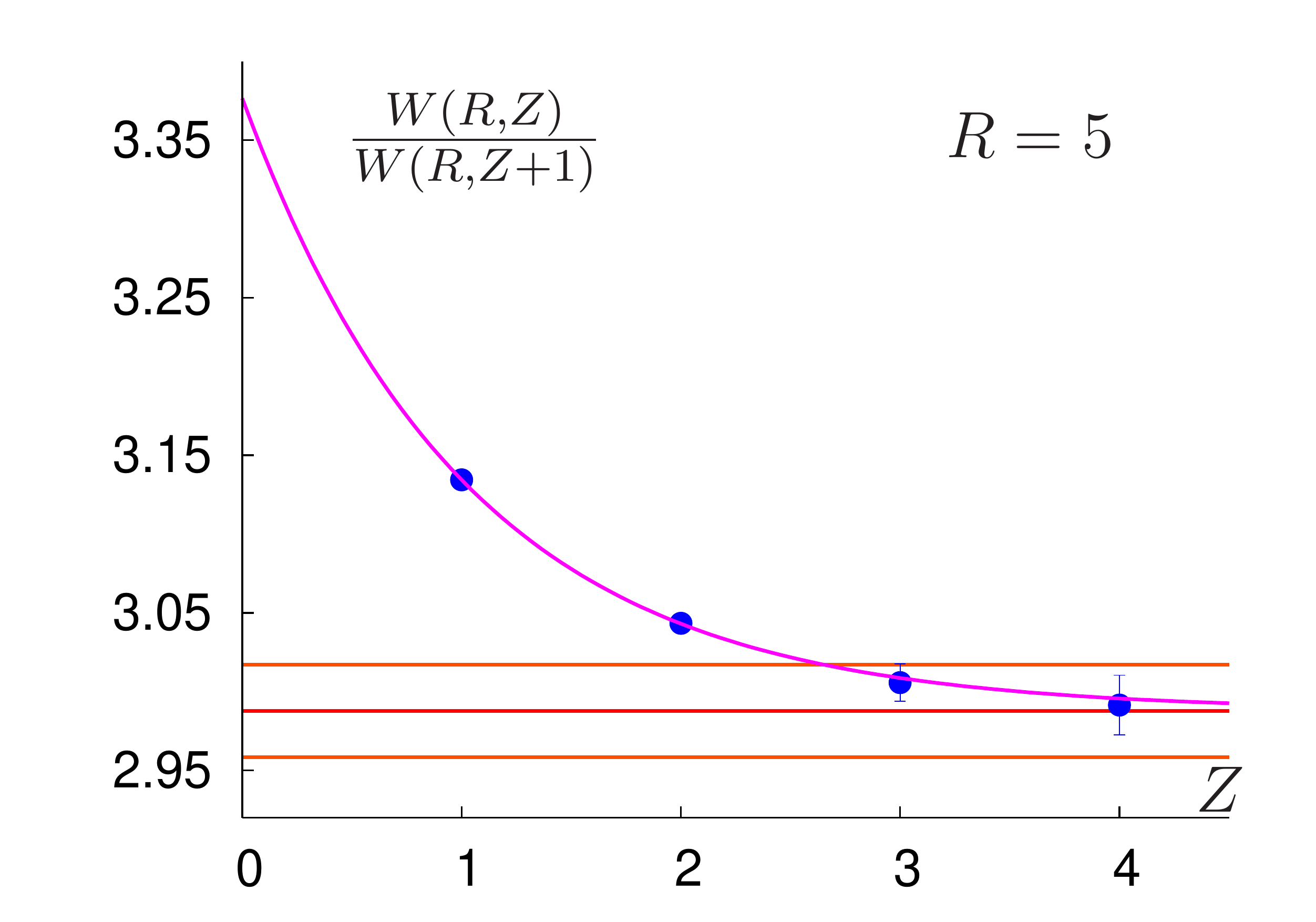}}
\end{center}
\caption{The ratio of spatial Wilson loops, for various $R$ at $\frac{T}{T_c} = 1.66$, on a $24^3 6$ lattice.}
\label{fig:stability}
\end{figure}

The spatial pseudo-potential is found from the ratio of spatial Wilson loops, $W(R,Z)$,

\begin{equation}
a V(R) = \lim_{Z \to \infty} \log \frac{W(R,Z)}{W(R,Z+1)}.
\end{equation}
It describes the interaction of a heavy quark and anti-quark separated by a distance $R$ in one spatial plane.  The "propagation" distance, $Z$, lies in the spatial direction perpendicular to the plane in which the sources lie.  The gauge fields were smeared 30 times to obtain a better projection onto the ground state.
In figure~\ref{fig:stability} we show an example of the ratios as function of $Z$.  To extract the pseudo-potentials we have fitted the data to the form equation~\ref{eqn:zextract}.  The horizontal lines denote the constant obtained from the fits.  As can be seen the fit results are close to the ratios at $Z = 3$.  We take the ratio at $Z=3$ to be sufficiently stable to calculate the potential.
\begin{equation}
\label{eqn:zextract}
\frac{W(R,Z)}{W(R,Z+1)} = c + a*exp(-bZ).
\end{equation}

\begin{figure}[ht]
\begin{center}
\includegraphics[angle=-90,totalheight=8.0cm]{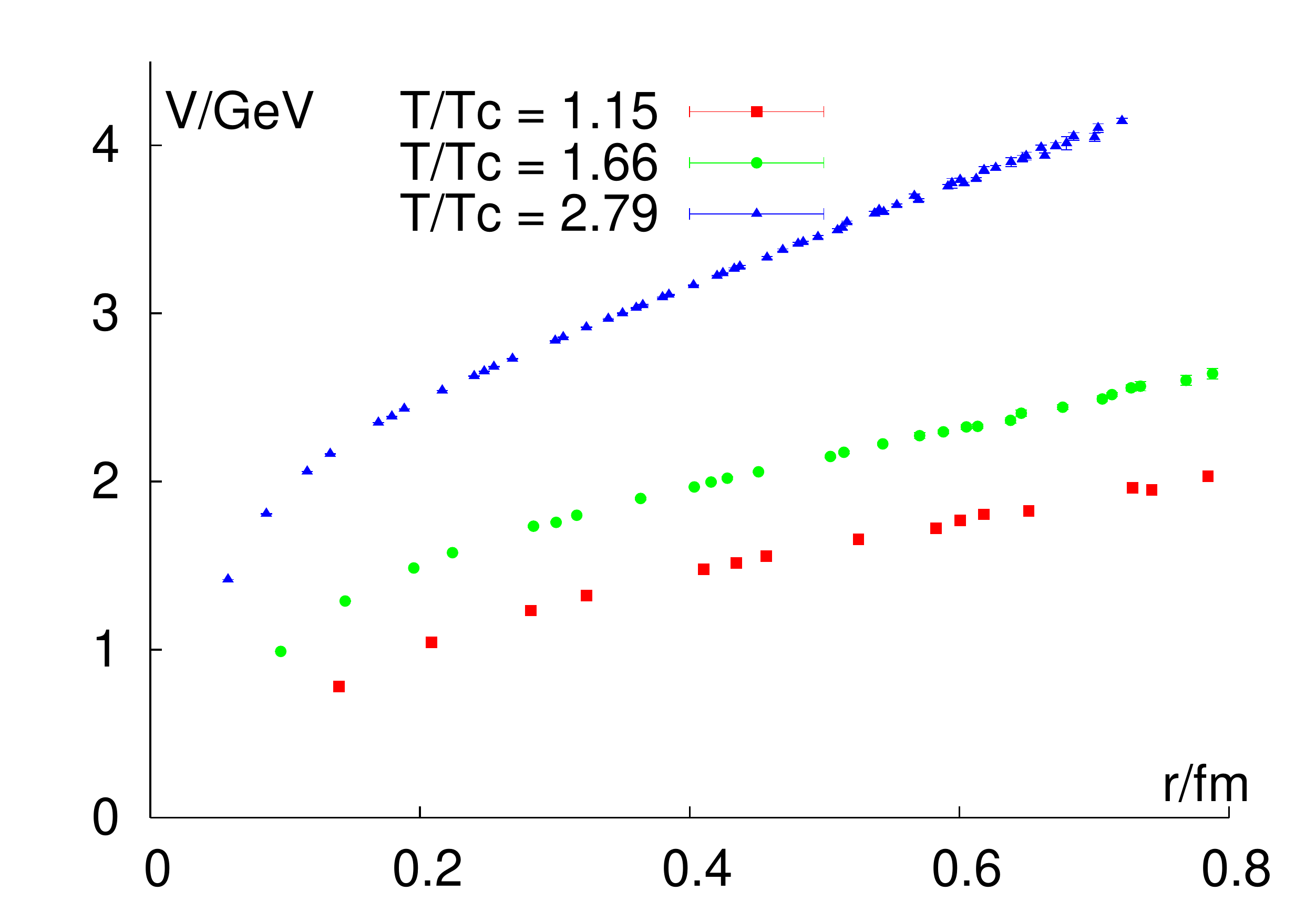}
\end{center}
\caption{Example of the spatial potential}
\label{fig:spatial_potential}
\end{figure}

In figure~\ref{fig:spatial_potential} the spatial potential, in physical units, is shown for 3 temperatures.  It can be readily seen, from the slope of the potentials at large separation that the spatial string tension increases with temperature.

In order to extract the spatial string tension we fit the pseudo-potential.  However, the choice of the fit ansatz is not unique.  At low temperatures the potential is well described by equation~\ref{eqn:3parampot},
\begin{equation}
\label{eqn:3parampot}
aV(R) = -\frac{\hat{\alpha}}{[R]} + \hat{c}_0 + \hat{\sigma}_s R,
\end{equation}
where the known cut-off effects are parametrised by replacing the distance R with an improved one.
\begin{equation}
\frac{1}{[R]} = 4 \pi \int_{-\pi}^{\pi} \frac{d^3 k}{(2 \pi)^3} e^{-k.R} D_{0}(k),
\end{equation}
with $D_{0}$ being the free lattice gluon propagator.  In this form we have a Coulombic $1/r$ term appropriate for a $3+1$ dimensional theory.  With increasing temperature, as the system becomes more and more $3$ dimensional, a logarithmic Coulombic term would be appropriate at small distances.  At the available distances we could not disentangle such a behaviour from a $1/r$ term coming from string fluctuations.  We did, however, observe that the coefficient of $1/r$ is decreasing with temperature.

Using the 3 parameter form for the potential careful judgement has to be made as to where to make minimum and maximum cuts.  The errors in the potential increase greatly with the separation $R$, as the signal seen in the ratio of the Wilson loops weakens.  The correlations in fluctuations in the potential also get stronger at large distances.  To control the correlations the fits to the potential are truncated at some maximum value, $\rmax$.  For the minimum value of the fit range we have generally taken values between $0.2fm$ and $0.3fm$.  This avoids the difficulties with a possible logarithmic term, yet, a $1/r$ contribution is still possible.  The fit range is typically $0.2fm \to 0.8fm$ for the temperatures studied.  Within this range the terms in the potential are all present such that the fits are stable against slight variations of $\rmin$.

\section{First approximation}
\begin{figure}[ht]
\begin{center}
\includegraphics[angle=-90,totalheight=9.2cm]{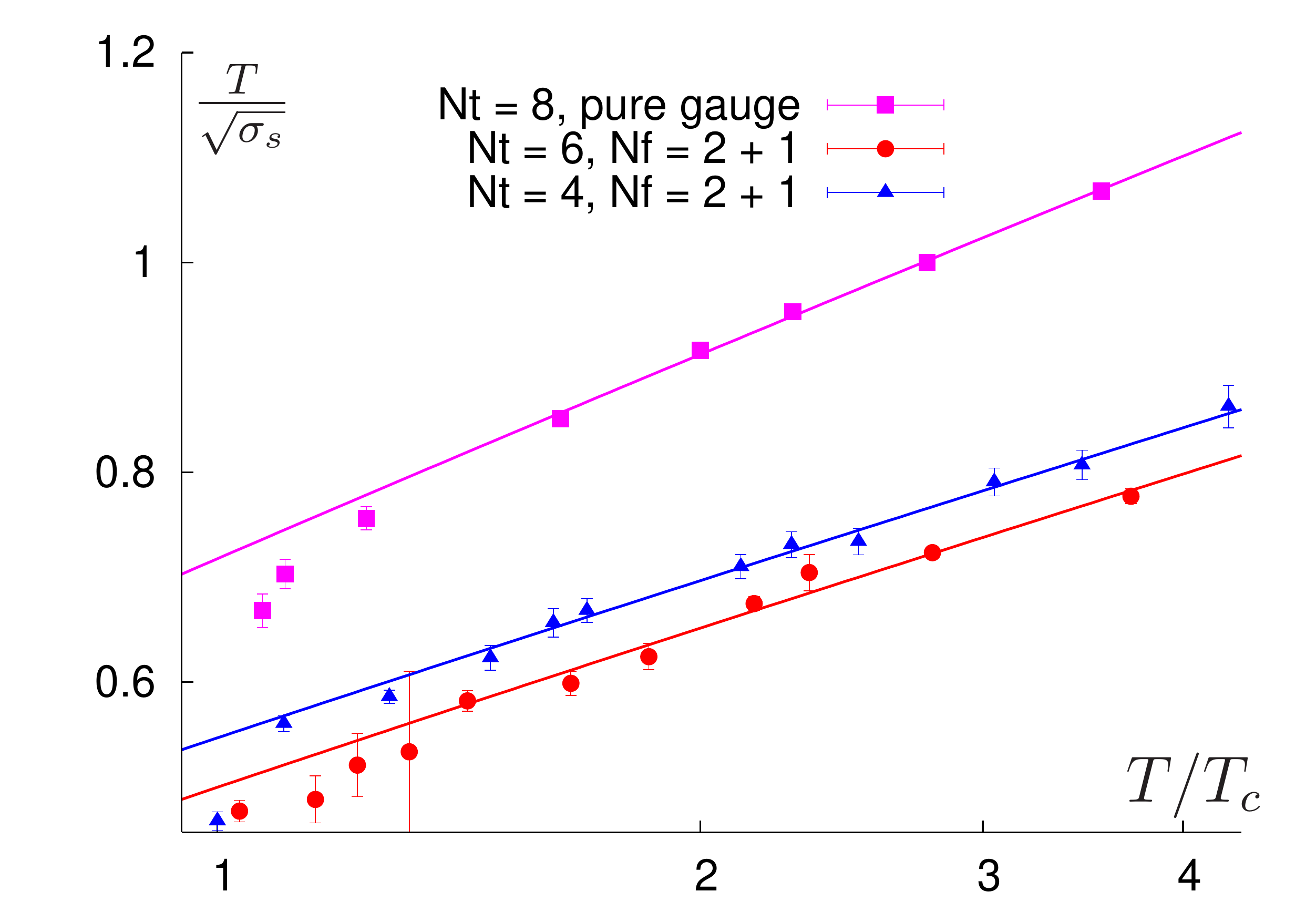}
\caption{Temperature dependence of the spatial string tension. The data has been fitted with equation \protect\ref{eqn:tform} and \protect\ref{eqn:2loop} for $T/T_c \geq 2$.}
\label{fig:0loop}
\end{center}
\end{figure}

We have seen that at high temperatures $3+1$ dimensional QCD may be reduced to a $3$ dimensional pure gauge theory.  We can now see what that implies for our measurements of the spatial string tension.  For the 3 dimensional gauge theory the coupling $g_M^2$ is dimensionful and sets a scale.

\begin{equation}
	\sqrt{\sigma_s} = c g^2_M
\end{equation}
From dimensional arguments we set this coupling to
\begin{equation}
g_M^2 = g^2(T) T.
\end{equation}
The spatial string tension is therefore obtained as
\begin{equation}
\frac{T}{\sqrt{\sigma_s}} = \frac{1}{c} g^{-2}(T).
\label{eqn:tform}
\end{equation}
Using a two-loop running coupling
\begin{equation}
g^{-2}(T) = 2 \beta_0 \ln \left( \frac{T}{\Lambda_\sigma} \right) + \frac{\beta_1}{\beta_0} \ln \left(2 \ln \left( \frac{T}{\Lambda_\sigma} \right)\right),
\label{eqn:2loop}
\end{equation}
with an unknown $\Lambda_\sigma$ parameter we arrive at a fit ansatz which has been fitted to our values of the spatial string tension for $T/T_c \geq 2$.  These fits can be seen in figure~\ref{fig:0loop} and in table~\ref{tab:0loop} the fit results are shown.

\begin{table}[ht]
\begin{center}
\begin{tabular}{|l|cc|}
\hline
& $\frac{\Lambda_\sigma}{T_c}$ & c \\
\hline
$N_\tau = 8$, Pure gauge& 0.099(9) & 0.575(14) \\
$N_\tau = 6$, $N_F = 2+1$ & 0.124(25) & 0.606(40) \\
$N_\tau = 4$, $N_F = 2+1$ & 0.100(23) & 0.607(39) \\
\hline
\end{tabular}
\end{center}
\caption{Fits parameters, $T > 2 T_c$}
\label{tab:0loop}
\end{table}

The values of the constant, $c$, in table~\ref{tab:0loop} can be compared with previous measurements in the 3 dimensional pure gauge theory~\cite{Teper:1998te,Lucini:2002wg,Karsch:1994af}.

\begin{equation}
c^\mathrm{gauge}_\mathrm{3D} = 0.553(1)~\footnote{The result from~\cite{Teper:1998te,Lucini:2002wg} is quoted}
\label{eqn:tepernum}
\end{equation}

The value for the $N_f = 2 + 1$ theory is in quite good agreement with that found in the three dimensional pure gauge theory, as well as four dimensional Yang-Mills theory, table~\ref{tab:0loop} and~\cite{Boyd:1996bx}.

\section{Two loop}
We have so far ignored perturbative effects in the relation between $g_E^2$ and $g^2(T)$, equation~\ref{eqn:0loopcoupling_electric}.
This relation is known at 2-loops~\cite{Laine:2005ai}.  Furthermore the connection between $g_M^2$ and $g_E^2$ is known to 2-loops~\cite{Giovannangeli:2003ti}.  Since the relationship between the spatial string tension and the coupling of MQCD has been accurately measured previously~\cite{Teper:1998te,Lucini:2002wg,Karsch:1994af}, equation~\ref{eqn:tepernum}, one can obtain a prediction of $\sigma_s$ of $3+1$ dimensional QCD.  This prediction requires a value for $\Lambda_{\msbar}$.

\begin{figure}[ht]
\begin{center}
\subfigure[$N_\tau = 4$]{\includegraphics[totalheight=4.5cm]{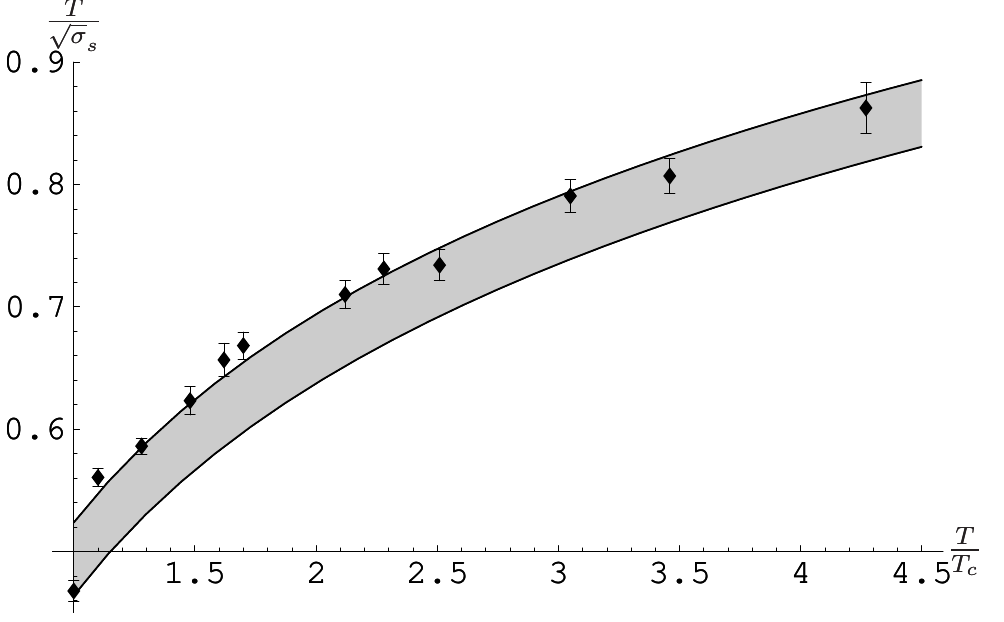}}
\subfigure[$N_\tau = 6$]{\includegraphics[totalheight=4.5cm]{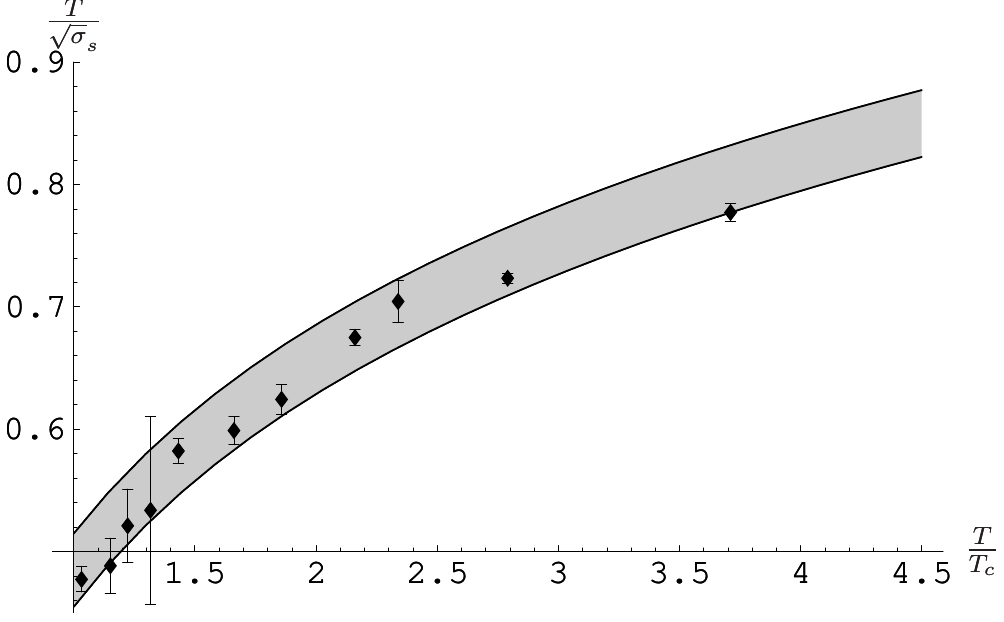}}
\end{center}
\label{fig:2loop_pred}
\caption{Comparison between lattice results and perturbative predictions.  The band represents the perturbative prediction with $0.260 < \Lambda_{\overline{MS}} < 0.341$, for both $N_\tau = 4$ and $N_\tau = 6$.}
\end{figure}

Our value of $\Lambda_{\overline{MS}}$ is obtained from a measurement of $\alpha_V(7.5 GeV) = 0.2082(40)$~\cite{Mason:2005zx}.  By matching the potential written in the $V$ scheme with the potential calculated in the $\overline{MS}$ scheme to two-loops~\cite{Schroder:1998vy}, we convert the value of $\alpha_V$ to $\alpha_{\overline{MS}}$.  $\Lambda_{\msbar}$ is then found using the two-loop beta function.  However there exists an ambiguity at which scale this procedure should be performed~\cite{Brodsky:1982gc}.  Matching $\alpha_V(7.5 GeV)$ with $\alpha_{\msbar}(7.5 GeV)$ gives $\Lambda_{\msbar} = 0.322(19)$, while matching it with $\alpha_{\msbar}(e^{-\frac{5}{6}} 7.5Gev)$ gives $\Lambda_{\msbar} = 0.285(15)$.  This uncertainty is taken into account in figure~\ref{fig:2loop_pred} where we compare the 2-loop perturbative prediction for the spatial string tension with the data.  It can be seen that it provides a good description of the data even down to low temperatures.

\section{Conclusions}
The spatial string tension, ${\sigma_s}$, was determined for $N_F = 2 + 1$ lattice QCD with lattice spacings corresponding to $N_\tau = 4,6$, for temperatures in the range $1 \leq \frac{T}{T_c} \leq 4$.  These string tensions were compared with predictions coming from dimensional reduction using the string tension of the $3$ dimensional pure gauge theory and the temperature dependent 2-loop coupling.

Although at present $\Lambda_{\msbar}$ is not yet available at high precision, the 2 loop reduction formulae provided a good description of the data, even at temperatures close to the transition temperature. 
\acknowledgments
This research was supported by the BMBF grant number O6BI401.

\end{document}